\documentclass[12pt]{article}

\begin{document}

\begin{center}
{\bf Note on Dirac$-$K\"{a}hler massless fields} \\
\vspace{5mm} S. I. Kruglov
\\
 \vspace{3mm} \textit{University of Toronto at Scarborough,\\ Physical
and Environmental Sciences Department, \\
1265 Military Trail, Toronto, Ontario, Canada M1C 1A4} \\

\vspace{5mm}
\end{center}

\begin{abstract}
We obtain the canonical and symmetrical Belinfante energy-momentum
tensors of Dirac$-$K\"{a}hler's fields. It is shown that the traces
of the energy-momentum tensors are not equal to zero. We find
the canonical and Belinfante dilatation currents which are not conserved, but
a new conserved dilatation current is obtained. It is pointed out that the conformal
symmetry is broken. The canonical quantization is performed and the propagator of the massless fields in the
first-order formalism is found.
\end{abstract}

\section{Introduction}

The Dirac$-$K\"{a}hler (DK) fields \cite{Ivanenko}, \cite{Kahler}
are paid much attention to due to the development of quantum
chromodynamics (QCD) on a lattice \cite{Becher}, \cite{Rabin},
\cite{Banks}, \cite{Joos}, \cite{Edwards}, \cite{Kanamori},
\cite{Jourjine}, \cite{Campos}. K\"{a}hler postulated an equation
in terms of inhomogeneous differential forms which is equivalent
to a set of antisymmetric tensor fields \cite{Ivanenko},
\cite{Kruglov}, \cite{Kruglov1}. In the matrix form the DK
equation for massive fields can be represented as a direct sum of
four Dirac equations. For massless fields, the DK equation comprises
the additional projection operator and the DK equation is not a sum of
Dirac equations.

In this paper, we investigate the massless DK boson fields. We
imply that masses of boson fields can appear due to the Higgs
mechanism. The relativistic wave equation describing massless DK
fields is the Dirac-like 16$\times$16 matrix equation with the
additional projection operator \cite{Kruglov}, \cite{Kruglov1}.
The Lagrangian for the DK massless boson fields possesses the
internal symmetry group $SO(3,1)$ \cite{Kruglov1}, and generators
of this group do not commute with the generators of the Lorentz
group.

We study here the dilatation symmetry of the massless DK fields.
The canonical, symmetrical Belinfante energy-momentum tensors, and
the dilatation current are obtained. The canonical quantization is
performed for the massless DK fields in the first-order formalism.

The paper is organized as follows. In Sec. 2, the massless DK field
equation is formulated in the matrix form. The canonical and the
symmetrical Belinfante energy-momentum tensors are found in Sec. 3.
We obtain the non-conserved canonical and Belinfante dilatation currents,
and a new conserved dilatation current is found. In Sec. 4, we consider
the canonical quantization of the DK massless fields and obtain the
matrix propagator. We discuss results obtained and possible applications of the
theory considered in Sec. 5.

The Heaviside units are chosen and the Euclidean metric is used, and $\hbar =c=1$.

\section{Dirac$-$K\"ahler equation for massless fields}

The massless DK fields obey the following tensor equations
\cite{Kruglov}, \cite{Kruglov1}:
\begin{equation}
\partial _\nu \varphi _{\mu \nu }-\partial _\mu \varphi
=0 ,\hspace{0.5in}\partial _\nu \widetilde{\varphi }_{\mu \nu
}-\partial _\mu \widetilde{\varphi } =0  ,\label{1}
\end{equation}
\begin{equation}
\partial _\mu \varphi _\mu =\varphi ,\hspace{0.3in}\partial _\mu \widetilde{
\varphi }_\mu =\widetilde{\varphi }, \label{2}
\end{equation}
\begin{equation}
\varphi _{\mu \nu }=\partial _\mu \varphi _\nu -\partial _\nu
\varphi _\mu -\varepsilon _{\mu \nu \alpha \beta }\partial _\alpha
\widetilde{\varphi } _\beta , \label{3}
\end{equation}
where the dual tensor is defined as
\begin{equation}
\widetilde{\varphi }_{\mu \nu }=\frac 12\varepsilon _{\mu \nu
\alpha \beta }\varphi _{\alpha \beta }  \label{4}
\end{equation}
and $\varepsilon _{\mu \nu \alpha \beta }$ is an antisymmetric
tensor with $\varepsilon _{1234}=-i$. We consider here the neutral
fields when the fields $\varphi $, $\widetilde{\varphi }$,
$(\varphi _m,\varphi _0) $, $(\widetilde{ \varphi }_m,\widetilde{
\varphi }_0) $ are real values. The generalization for charged
fields is straightforward: the fields become complex values. One
may introduce the electric and magnetic sources in Eq. (1)
\cite{Kruglov1}. Then, we have the Maxwell equations with electric
and magnetic charges in the dual-symmetric form. The fields
$\varphi $, $\widetilde{\varphi }$ play the role of the general
gauge \cite{Faddeev}. To have the Lorentz gauge, one can put
$\varphi =\widetilde{\varphi }=0$.

Equations (1)$-$(3) can be represented in the form of the Dirac-like
equation with $16\times 16$ dimensional Dirac matrices
\cite{Kruglov}, \cite{Kruglov1}. It should be noted that the
fields $\varphi $, $\widetilde{\varphi }$, $\varphi _\mu$,
$\widetilde{ \varphi }_\mu$ have different dimensions. Thus,
in order to formulate the first-order equations, one needs to
introduce the dimensional parameter. We use the notations $\psi
_0=-\varphi $, $\psi _\mu =\kappa\varphi _\mu $, $\psi _{[\mu \nu
]}=\varphi _{\mu \nu },$ $\widetilde{\psi }_\mu
=i\kappa\widetilde{ \varphi }_\mu ,$ $\widetilde{\psi
}_0=-i\widetilde{\varphi },$ $e_{\mu \nu \alpha \beta
}=i\varepsilon _{\mu \nu \alpha \beta }$ ($e_{1234}=1$), where the
parameter $\kappa$ has the dimension of the mass\footnote{In
\cite{Kruglov}, \cite{Kruglov1}, we have used instead of $\kappa$
the parameter $m_2$; thus, $\kappa=m_2$.}. With these notations,
equations (1)$-$(3) may be rewritten as
\[
\partial _\mu \psi _\mu +\kappa\psi _0=0 ,~~~~
\partial_\mu \widetilde{\psi }_\mu +\kappa\widetilde{\psi }_0=0 ,
\]
\begin{equation}
\partial_\nu \psi _{[\mu \nu ]}+\partial _\mu \psi _0=0 ,~~~~
\partial_\nu i\widetilde{\psi}_{[\mu \nu ]}+\partial _\mu \widetilde{\psi}_0=0 ,   \label{5}
\end{equation}
\[
\partial_\nu \psi_\mu -\partial_\mu \psi _\nu -e_{\mu \nu \alpha \beta
}\partial_\alpha \widetilde{\psi }_\beta +\kappa\psi_{[\mu \nu
]}=0 ,
\]
Introducing the 16-component wave function
\begin{equation}
\Psi (x)=\left\{ \psi _A(x)\right\} =\left(
\begin{array}{c}
\psi_0(x)\\
\psi _\mu (x)\\ \psi_{[\mu\nu]}(x)\\\widetilde{\psi}_\mu (x)
\\ \widetilde{\psi}_0 (x)
\end{array}
\right), \label{6}
\end{equation}
where $A=0,\mu ,[\mu \nu ], \widetilde{\mu },\widetilde{0}$; $\psi
_{\widetilde{\mu}}\equiv\widetilde{\psi}_\mu$, $\psi
_{\widetilde{0}}\equiv\widetilde{\psi}_0$, (5) can be cast in
the form of the first-order wave equation \cite{Kruglov},
\cite{Kruglov1}:
\begin{equation}
\left( \Gamma _\nu \partial _\nu +\kappa P\right) \Psi (x)=0.
\label{7}
\end{equation}
The $16\times 16 $ matrices $\Gamma _\nu $ are given by
\[
\Gamma _\nu =\beta _\nu ^{(+)}+\beta _\nu ^{(-)} ,
\hspace{0.3in}\beta _\nu ^{(+)}=\beta _\nu ^{(1)}+\beta _\nu
^{(\widetilde{0})} ,\hspace{0.3in}\beta _\nu ^{(-)}=\beta _\nu
^{(\widetilde{1})}+\beta _\nu ^{(0)} ,
\]
\begin{equation}
\beta _\nu ^{(1)}=\varepsilon ^{\mu ,[\mu \nu ]}+\varepsilon
^{[\mu \nu ],\mu } ,\hspace{0.3in}\beta _\nu
^{(\widetilde{1})}=\frac 12e_{\mu \nu \rho \omega }\left(
\varepsilon ^{\widetilde{\mu },[\rho \omega ]}+\varepsilon ^{[\rho
\omega ],\widetilde{\mu }}\right) , \label{8}
\end{equation}
\[
\beta _\nu ^{(\widetilde{0})}=\varepsilon ^{\widetilde{\nu
},\widetilde{0} }+\varepsilon ^{\widetilde{0},\widetilde{\nu }} ,
\hspace{0.3in}\beta _\nu ^{(0)}=\varepsilon ^{\nu ,0}+\varepsilon
^{0,\nu } .
\]
Equation (7) describes the massless DK fields and it is not the direct
sum of four Dirac equations because of the presence of the projection
operator $P$. In the case of massive fields, the parameter
$\kappa$ is replaced by the mass $m$, and the projection operator
$P$ is replaced by the unit 16$\times$16 matrix. As a result the equation for
massive DK fields represents the sum of four Dirac equations. But
the Lorentz transformations for bosonic fields mix ``flavors"
\cite{Kruglov}, \cite{Kruglov1}. Matrices $\beta _\nu ^{(1)}$,
$\beta _\nu ^{(\widetilde{1})}$ and $\beta _\nu ^{(0)}$, $\beta
_\nu ^{(\widetilde{0})}$ obey the Petiau$-$Duffin$-$Kemmer algebra,
and the $\Gamma _\nu $ are the $16\times 16 $ Dirac-like matrices:
\begin{equation}
\Gamma _\nu \Gamma _\mu +\Gamma _\mu \Gamma _\nu =2\delta _{\mu
\nu } .\label{9}
\end{equation}
We have explored here the matrices $\varepsilon ^{A,B}$ with the
properties: $\varepsilon ^{A,B}\varepsilon ^{C,D}=\varepsilon
^{A,D}\delta _{BC}$, $\left( \varepsilon ^{A,B}\right)
_{CD}=\delta _{AC}\delta _{BD}$ and the indexes run
$A,B,C,D=1,2,...,16$. The matrix $P$ is the projection matrix,
$P^2=P$, and is given by
\begin{equation}
P=\varepsilon ^{0,0}+\frac 12\varepsilon ^{[\mu \nu ],[\mu \nu
]}+\varepsilon ^{\widetilde{0},\widetilde{0}}. \label{10}
\end{equation}

Now we consider the form-invariance of (7) under the Lorentz transformations.
Coordinate transformations read as follows:
\begin{equation}
x_\mu ^{\prime }=L_{\mu \nu }x_\nu ^{\prime } , \label{11}
\end{equation}
where the Lorentz matrix $L=\{L_{\mu \nu }\}$ has the properties:
$L_{\mu \alpha }L_{\nu \alpha }=\delta _{\mu \nu }$.
The Lorentz transformations of coordinates (11) generate the wave function
transformations
\begin{equation}
\Psi ^{\prime }(x^{\prime })=T\Psi (x) , \label{12}
\end{equation}
with the $16\times 16$ matrix $T$. Then the first-order equation (7) becomes
\begin{equation}
\left( \Gamma _\mu \partial _\mu ^{\prime }+\kappa P\right) \Psi ^{\prime
}(x^{\prime })=\left( \Gamma _\mu L_{\mu \nu }\partial _\nu
+\kappa P\right) T\Psi (x)=0 , \label{13}
\end{equation}
where $\partial _\mu ^{\prime }=L_{\mu \nu}\partial _\nu $.
Equation (7) is form-invariant under the Lorentz transformations
if the equations
\begin{equation}
\Gamma _\mu TL_{\mu \nu }=T\Gamma _\nu,~~~~~~PT=TP  \label{14}
\end{equation}
hold. The matrix $T$ for the finite Lorentz transformations is given by
\begin{equation}
T=\exp \left( \frac 12\varepsilon _{\mu \nu }J_{\mu \nu }\right) ,
\label{15}
\end{equation}
where $J_{\mu \nu}$ are the generators of the Lorentz group transformations.
One can verify that the matrix $T$ with the generators for bosonic fields
\begin{equation}
J_{\mu \nu }=\frac 14\left( \Gamma _\mu \Gamma _\nu -\Gamma _\nu
\Gamma _\mu +\overline{\Gamma }_\mu \overline{\Gamma }_\nu
-\overline{\Gamma }_\nu \overline{\Gamma }_\mu \right) ,
\label{16}
\end{equation}
obeys (14). We have introduced
the matrices $\overline{\Gamma }_\nu $, which satisfy the Dirac
algebra and are given by
\begin{equation}
\overline{\Gamma }_\nu =\beta _\nu ^{(+)}-\beta _\nu ^{(-)} .
\label{17}
\end{equation}
The matrices $\overline{\Gamma } _\nu $ commute with $\Gamma _\mu
$: $ \Gamma _\mu \overline{\Gamma }_\nu  =\overline{\Gamma }_\nu
\Gamma _\mu $.
At the infinitesimal Lorentz transformations (14) become
\begin{equation}
\Gamma _\mu J_{\alpha \nu }-J_{\alpha \nu }\Gamma _\mu =\delta
_{\alpha \mu }\Gamma _\nu -\delta _{\nu \mu }\Gamma _\alpha, ~~~~PJ_{\mu\nu}=J_{\mu\nu}P .
\label{18}
\end{equation}
The generators (16) obey (18).

The Lorentz-invariant is $\overline{\Psi }\Psi
=\Psi ^{+}\eta \Psi$ ($\Psi ^{+}$ is the Hermitian-conjugate wave
function), where the Hermitian matrix, $\eta $ is
\begin{equation}
\eta =\Gamma _4\overline{\Gamma }_4 . \label{19}
\end{equation}
From (8),(17), we obtain
\begin{equation}
\eta =-\varepsilon ^{0,0}+\varepsilon ^{m,m}-\varepsilon ^{4,4}
+\varepsilon ^{[m4],[m4]}-\frac{1}{2}\varepsilon ^{[mn],[mn]}+
\varepsilon ^{\widetilde{0},\widetilde{0}} +\varepsilon
^{\widetilde{4},\widetilde{4}}-\varepsilon
^{\widetilde{m},\widetilde{m}}. \label{20}
\end{equation}
Taking into consideration that the fields $\varphi $,
$\widetilde{\varphi }$, $(\varphi _m,\varphi _0) $, $(\widetilde{
\varphi }_m,\widetilde{ \varphi }_0) $ are real values, we find
from (20) the ``conjugated" function:
\begin{equation}
\overline{\Psi}(x)=\Psi
^{+}(x)\eta=\left(-\psi_0(x),\psi_\mu(x),-\psi_{[\mu\nu]}(x),
\widetilde{\psi}_\mu(x),-\widetilde{\psi}_0(x)\right) \label{21}
\end{equation}
which obeys the equation
\begin{equation}
\overline{ \Psi }(x)\left( \Gamma _\mu
\overleftarrow{\partial}_\mu -\kappa P\right) =0  .\label{22}
\end{equation}
It follows from (7),(22) that the electric current
\begin{equation}
J_\mu (x)=i\overline{ \Psi }(x)\Gamma_\mu\Psi(x)\label{23}
\end{equation}
is conserved: $\partial_\mu J_\mu (x)=0$. In addition, one may
verify with the help of (6),(8),(21), that for the real fields,
it vanishes, $J_\mu (x)=0$, as the fields are neutral.

Solutions to (7) with definite energy and momentum are given by
\begin{equation}
\Psi_s^{(\pm)}(x)=\sqrt{\frac{\kappa}{2k_0 V}}u_s(\pm k)\exp(\pm
ikx) , \label{24}
\end{equation}
where $V$ is the normalization volume, $k^2=
\textbf{k}^2-k_0^2=0$, and $s$ is the spin index which corresponds
to scalar, vector, pseudovector and pseudoscalar states and runs
eight values: $s=0,n,\widetilde{n},\widetilde{0}$. The $u_s(\pm
k)$ obeys the equation for the field function in momentum
space:
\begin{equation}
\left(\pm i\widehat{k}+\kappa P\right)u_s(\pm k)=0 ,  \label{25}
\end{equation}
where $\widehat{k}=\Gamma _\mu k_\mu $. One found in
\cite{Kruglov}, \cite{Kruglov1} the minimal equation for the
matrix of (25) $B_\pm=\pm i\widehat{k}+\kappa P$:
\begin{equation}
B_\pm\left( B_\pm-\kappa\right) =0 , \label{26}
\end{equation}
so that the projection operator ($\alpha_\pm =\alpha_\pm^2$)
extracting solutions to (25) is given by
\begin{equation}
\alpha_\pm =\frac{\kappa-B_\pm}{\kappa}=\overline{P}\mp
\frac{i\widehat{k}}{\kappa}, \label{27}
\end{equation}
and $\overline{P}$ is the projection operator:
\begin{equation}
\overline{P}=1-P=\varepsilon ^{\mu ,\mu }+\varepsilon
^{\widetilde{\mu }, \widetilde{\mu }} , \label{28}
\end{equation}
and $\overline{P}P=P\overline{P}=0$. Every column of the matrix
$\alpha$ is the solution to (25). One can find the projection
matrix-dyads, extracting solutions with different spins and spin
projections in \cite{Kruglov}, \cite{Kruglov1}. From (23),(25),
and the condition $J_\mu (x)= 0$ for the neutral fields, one
obtains
\begin{equation} \overline{u}_s(\pm
k)\hat{k} u_{s}(\pm k)=0,~~~~\overline{u}_s(\pm k)P u_{s}(\pm k)=0
. \label{29}
\end{equation}
Equation (27) will be used in the second quantization theory for
obtaining the propagator of DK fields.

\section{The energy-momentum tensor}

The Lagrangian of massless DK fields in the first-order formalism
can be written as
\begin{equation}
\mathcal{L}=-\frac{1}{2}\overline{\Psi }(x)\left( \Gamma _\mu
\partial _\mu +\kappa P\right) \Psi (x)+\frac{1}{2}\overline{\Psi }(x)\left( \Gamma _\mu
\overleftarrow{\partial _\mu} -\kappa P\right) \Psi (x) .
\label{30}
\end{equation}
Equations (7),(22) follow from Lagrangian (30) by varying the
corresponding action on the wave functions $\overline{\Psi }(x)$,
$\Psi (x)$. For the neutral DK fields, the Lagrangian (30) reduces
to
\begin{equation}
\mathcal{L}=-\overline{\Psi }(x)\left( \Gamma _\mu
\partial _\mu +\kappa P\right) \Psi (x).
\label{31}
\end{equation}
With the help of (6),(8),(21), Lagrangian (31) becomes
\[
{\cal L}=\psi_0 \partial _\mu\psi_\mu -\psi_\mu\partial _\mu
\psi_0 -\psi_\rho\partial _\mu\psi_{[\rho\mu]}+
\psi_{[\rho\mu]}\partial _\mu\psi_\rho+{}^\star
\psi_{[\rho\mu]}\partial_\mu \widetilde{\psi}_\rho
\]
\vspace{-8mm}
\begin{equation}
\label{32}
\end{equation}
\vspace{-8mm}
\[
- \widetilde{\psi}_\rho\partial_\mu{}^\star
\psi_{[\rho\mu]}-\widetilde{\psi}_\mu\partial_\mu
\widetilde{\psi}_0+ \widetilde{\psi}_0\partial_\mu
\widetilde{\psi}_\mu+\kappa\left(\psi_0^2
+\frac{1}{2}\psi_{[\rho\mu]}^2+\widetilde{\psi}_0^2\right),
\]
where
\begin{equation}
{}^\star \psi_{[\mu\nu]}=\frac 12 e _{\mu \nu \alpha \beta }\psi
_{[\alpha \beta] }=i\widetilde{\psi}_{[\mu\nu]}, \label{33}
\end{equation}
$e _{\mu \nu \alpha \beta }$ is antisymmetric tensor ($e _{1234
}=1$). It should be noted that ${}^\star \psi_{[\mu\nu]}$ is not a
dual tensor because $e _{\mu \nu \alpha \beta }=i\varepsilon _{\mu
\nu \alpha \beta }$, and the dual tensor is defined by Eq.(4). One
can verify that the Euler$-$Lagrange equations
\begin{equation}
\frac{\partial{\cal
L}}{\partial\psi_A}-\partial_\mu\frac{\partial{\cal
L}}{\partial(\partial_\mu\psi_A)}=0 \label{34}
\end{equation}
with ${\cal L}$ given by (32), where $A=0,\mu ,[\mu \nu ],
\widetilde{\mu },\widetilde{0}$, lead to (5). For fields
obeying the equations of motion (5) (or (7) and (22)) the Lagrangians
(30),(31) and (32) vanish similarly to the Dirac Lagrangian.

The canonical energy-momentum tensor in the first-order formalism
is given by
\begin{equation}
T^c_{\mu\nu}=\frac{\partial{\cal
L}}{\partial(\partial_\mu\Psi(x))}\partial_\nu\Psi(x)-\delta_{\mu\nu}{\cal
L},  \label{35}
\end{equation}
and using Eq.(31) it becomes
\begin{equation}
T^c_{\mu\nu}=\left(\partial_\nu \overline{\Psi}
(x)\right)\Gamma_\mu \Psi (x). \label{36}
\end{equation}
We took into consideration here that for fields obeying equations
of motion ${\cal L}=0$. One obtains from (6),(8),(21), and (36)
the expression in the tensor form:
\[
T^c_{\mu\nu}=\psi_0\partial_\nu \psi_\mu-\psi_\mu\partial_\nu
\psi_0-\psi_\rho\partial_\nu \psi_{[\rho\mu]} +\psi_{[\rho\mu]}
\partial_\nu \psi_\rho+{}^\star
\psi_{[\rho\mu]}\partial_\nu \widetilde{\psi}_\rho
\]
\vspace{-8mm}
\begin{equation}
\label{37}
\end{equation}
\vspace{-8mm}
\[
- \widetilde{\psi}_\rho\partial_\nu{}^\star
\psi_{[\rho\mu]}-\widetilde{\psi}_\mu\partial_\nu
\widetilde{\psi}_0+ \widetilde{\psi}_0\partial_\nu
\widetilde{\psi}_\mu.
\]
It follows from the field equations that the energy-momentum tensor
(37) (and (36)) is conserved tensor, $\partial_\mu
T^c_{\mu\nu}=0$. Contrary to classical electrodynamics
\cite{Landau} the energy-momentum tensor (37) is not the symmetric
tensor, $T^c_{\mu\nu}\neq T^c_{\nu\mu}$.

Now, we investigate the dilatation symmetry \cite{Coleman}. The
canonical dilatation current in the first-order formalism is given
by
\begin{equation}
D_\mu^c=x_\nu T_{\mu\nu}^{c}+\Pi_\mu \Psi, \label{38}
\end{equation}
where
\begin{equation}
\Pi_\mu=\frac{{\partial\cal
L}}{\partial\left(\partial_\mu\Psi\right)} =-\overline{\Psi
}\Gamma_\mu. \label{39}
\end{equation}
We note that for the bosonic fields, the matrix $\textbf{d}$ in
\cite{Coleman} defining the field dimension, is the unit matrix.
As the electric current $J_\mu=i\overline{\Psi}\Gamma_\mu\Psi$,
for neutral fields equals $0$, the last term in Eq.(38) vanishes.
As a result, we obtain non-zero divergence of the canonical
dilatation current
\begin{equation}
\partial_\mu D_\mu^c=T_{\mu\mu}^{c}=-\kappa\left(\psi_0^2(x) +
\frac{1}{2}\psi_{[\mu\nu]}^2(x)+\widetilde{\psi}_0^2(x)\right).
\label{40}
\end{equation}
The dilatation current $D_\mu^c$ is not conserved current. The similar expression was found in the
first-order formulation of generalized electrodynamics with an
additional scalar field \cite{Kruglov2}. Later, we will obtain
new conserved current. Expression (40) also can
be obtained from the relationship \cite{Coleman}
\[
\partial_\mu D_\mu^c=2\Pi_\mu\partial_\mu\Psi +\frac{\partial{\cal L}}{\partial\Psi}\Psi -4{\cal L}
=\kappa\overline{\Psi }(x)P\Psi(x)
\]
\vspace{-8mm}
\begin{equation}
\label{41}
\end{equation}
\vspace{-8mm}
\[
=-\kappa\left(\psi_0^2(x) +
\frac{1}{2}\psi_{[\mu\nu]}^2(x)+\widetilde{\psi}_0^2(x)\right).
\]
It should be noted that for the plane-wave solution (24), we find
from Eq.(5)
\begin{equation}
\kappa\psi_{[\mu \nu ]}(k)=k_\mu \psi _\nu (k) -k_\nu \psi_\mu (k)
+e_{\mu \nu \alpha \beta }k_\alpha \widetilde{\psi }_\beta (k) ,
\label{42}
\end{equation}
and
\begin{equation}
\frac{1}{2}\psi^2_{[\mu \nu
]}(k)=-\psi_0^2(k)-\widetilde{\psi}_0^2(k). \label{43}
\end{equation}
As a result, the right sides of (40),(41) vanish and the
dilatation current is conserved, $\partial_\mu D_\mu^c=0$. But, in
the general configuration of fields, the dilatation current $D_\mu^c$ is
not conserved.

Now, we find the symmetrical energy-momentum tensor. The general
expression for the symmetrical Belinfante energy-momentum tensor is
given by (see \cite{Coleman}):
 \begin{equation}
T_{\mu\nu}^{B}=T_{\mu\nu}^{c}+ \partial_\beta X_{\beta\mu\nu},
\label{44}
\end{equation}
and
\begin{equation}
X_{\beta\mu\nu}=\frac{1}{2}\left[\Pi_\beta J_{\mu\nu}\Psi- \Pi_\mu
J_{\beta\nu}\Psi-\Pi_\nu J_{\beta\mu}\Psi\right]. \label{45}
\end{equation}
The tensor (45) is antisymmetrical in indexes $\beta,\mu$, and
therefore $\partial_\mu\partial_\beta X_{\beta\mu\nu}=0$. As a
result, $\partial_\mu T_{\mu\nu}^{B}=\partial_\mu
T_{\mu\nu}^{c}=0$. After some calculations, we obtain from (16)
the expression for generators of the Lorentz group
\begin{equation}
J_{\mu\nu}=\varepsilon^{\mu,\nu}-\varepsilon^{\nu,\mu}
+\varepsilon^{[\lambda\mu],[\lambda\nu]}-\varepsilon^{[\lambda\nu],[\lambda\mu]}
+\varepsilon^{\widetilde{\mu},\widetilde{\nu}}-\varepsilon^{\widetilde{\nu},\widetilde{\mu}}.
\label{46}
\end{equation}
Replacing expression (46) in Eq.(45), and taking into account
(39), after calculations, we obtain
\[
X_{\alpha\mu\nu}=\delta_{\mu\nu}\psi_0\psi_\alpha-\delta_{\alpha\nu}\psi_0\psi_\mu+
\delta_{\alpha\nu}\psi_\lambda\psi_{[\lambda\mu]}-\delta_{\mu\nu}\psi_\lambda\psi_{[\lambda\alpha]}
+2\psi_\nu\psi_{[\mu\alpha]}
\]
\vspace{-8mm}
\begin{equation}
\label{47}
\end{equation}
\vspace{-8mm}
\[
+\delta_{\mu\nu}\widetilde{\psi}_0\widetilde{\psi}_\alpha
-\delta_{\alpha\nu}\widetilde{\psi}_0\widetilde{\psi}_\mu+
\widetilde{\psi}_\nu{}^\star\psi_{[\mu\alpha]}+
e_{\beta\alpha\mu\lambda}\widetilde{\psi}_\beta\psi_{[\lambda\nu]}.
\]
From (37),(44),(47), we obtain the Belinfante energy-momentum
tensor
\[
T_{\mu\nu}^{B}=-2\psi_\mu\partial_\nu\psi_0-2\widetilde{\psi}_\mu\partial_\nu\widetilde{\psi}_0
+2\psi_{[\lambda\mu]}\partial_\nu\psi_\lambda-\widetilde{\psi}_\beta\partial_\nu{}^\star\psi_{[\beta\mu]}
+{}^\star\psi_{[\beta\mu]}\partial_\nu\widetilde{\psi}_\beta
\]
\vspace{-7mm}
\begin{equation}
\label{48}
\end{equation}
\vspace{-7mm}
\[
+\partial_\alpha\left[2\psi_\nu\psi_{[\mu\alpha]}+\widetilde{\psi}_\nu{}^\star\psi_{[\mu\alpha]}
+e_{\beta\alpha\mu\lambda]}\widetilde{\psi}_\beta\psi_{[\lambda\nu]}
+\delta_{\mu\nu}\left(\psi_0\psi_\alpha+\widetilde{\psi}_0\widetilde{\psi}_\alpha
-\psi_\lambda\psi_{[\lambda\alpha]}\right)\right].
\]
Expression (48) with the help of equations of motion (5) can be
represented in the symmetrical form
\[
T_{\mu\nu}^{B}=-2\psi_\mu\partial_\nu\psi_0-2\psi_\nu\partial_\mu\psi_0
-2\widetilde{\psi}_\mu\partial_\nu\widetilde{\psi}_0
-2\widetilde{\psi}_\nu\partial_\mu\widetilde{\psi}_0 +
\kappa\biggl(2F_{\mu\alpha}F_{\alpha\nu}+G_{\mu\alpha}G_{\alpha\nu}
\]
\vspace{-7mm}
\begin{equation}
\label{49}
\end{equation}
\vspace{-7mm}
\[
-{}^\star G_{\mu\alpha}{}^\star G_{\alpha\nu}+{}^\star
F_{\mu\alpha}G_{\alpha\nu}+{}^\star
G_{\mu\alpha}F_{\alpha\nu}\biggr)
+\delta_{\mu\nu}\left[\partial_\alpha\left(\psi_0\psi_\alpha+\widetilde{\psi}_0\widetilde{\psi}_\alpha
-\psi_\lambda\psi_{[\lambda\alpha]}\right)+\widetilde{\psi}_\alpha\partial_\alpha\widetilde{\psi}_0\right],
\]
where we use the notation
\begin{equation}
\kappa F_{\mu\nu}=\partial_\mu\psi_\nu -\partial_\nu\psi_\mu,~~~~
\kappa G_{\mu\nu}=\partial_\mu\widetilde{\psi}_\nu
-\partial_\nu\widetilde{\psi}_\mu. \label{50}
\end{equation}
It is easy to find, with the help of field equations (5), the
trace of the Belinfante energy-momentum tensor (49):
\begin{equation}
T_{\mu\mu}^{B}=4\partial_\mu\left(\psi_0
\psi_\mu+\widetilde{\psi}_0\widetilde{\psi}_\mu \right).
\label{51}
\end{equation}

 The modified dilatation current is given by \cite{Coleman}
 \begin{equation}
D_\mu^B=x_\alpha T_{\mu\alpha}^{B}+V_\mu, \label{52}
\end{equation}
where the field-virial $V_\mu$ is defined as
 \[
V_\mu=\Pi_\mu \Psi-\Pi_\alpha J_{\alpha\mu}\Psi=\overline{\Psi
}\Gamma_\alpha J_{\alpha\mu}\Psi
\]
\vspace{-7mm}
\begin{equation}
\label{53}
\end{equation}
\vspace{-7mm}
\[
=\psi_\lambda\psi_{[\lambda\mu]}+
\widetilde{\psi}_\lambda{}^\star\psi_{[\lambda\mu]}
-3\left(\psi_0\psi_\mu+\widetilde{\psi}_0\widetilde{\psi}_\mu\right).
\]
It is easy to verify that $X_{\alpha\mu\mu}=-V_\alpha$. As a
result, the divergence of the Belinfante dilatation current
becomes
\begin{equation}
\partial_\mu D_\mu^B=T_{\mu\mu}^{B}+\partial_\mu V_\mu=
\partial_\mu D_\mu^c=T_{\mu\mu}^c. \label{54}
\end{equation}
The divergences of the Belinfante and canonical
dilatation currents are the same. Thus, the currents $D_\mu^c$, $D_\mu^B$ are not
conserved, but because the trace of the Belinfante energy-momentum tensor (51) is a total divergence,
we can introduce a new conserved
current\footnote{I am grateful to Yu Nakayama for his remarks.}:
\begin{equation}
D_\mu=x_\alpha T_{\mu\alpha}^{B}-4\left(\psi_0
\psi_\mu+\widetilde{\psi}_0\widetilde{\psi}_\mu\right), \label{55}
\end{equation}
so that $\partial_\mu D_\mu=0$. Thus, massless DK fields possess the dilatation symmetry
with new dilatation current (55). The similar conserved
current can be introduced also for generalized electrodynamics
\cite{Kruglov2}, \cite{Kruglov3}. It should be noted that the conformal invariance is broken because the field-virial
$V_\mu$ is not a total derivative of some local quantity \cite{Coleman}.

\section{Canonical quantization}

To perform the canonical quantization, we define the momenta in
the matrix form from Eq.(31):
\begin{equation}
\pi (x)=\frac{\partial\mathcal{L}}{\partial(\partial_0\Psi
(x))}=i\overline{\Psi}\Gamma_4.
 \label{56}
\end{equation}
Then, using the quantum commutator
\[
[\Psi_M(\textbf{x},t),\pi_N (\textbf{y},t)]= i \delta_{MN}
\delta(\textbf{x}-\textbf{y}),
\]
we obtain from (56) the commutation relation
\begin{equation}
\left[\Psi_M(\textbf{x},t),
\left(\overline{\Psi}(\textbf{y},t)\Gamma_4 \right)_N\right ]=
\delta_{MN}\delta(\textbf{x}-\textbf{y}) .\label{57}
\end{equation}
Equation (57), with the help of (6),(8),(21), leads to simultaneous
field commutators for the components $\psi_A (x)$:
\[
\left[\psi_0(\textbf{x},t),\psi_4(\textbf{y},t)\right ]=
\delta(\textbf{x}-\textbf{y}),~~~~\left[\widetilde{\psi}_0(\textbf{x},t),
\widetilde{\psi}_4(\textbf{y},t)\right]=
\delta(\textbf{x}-\textbf{y}),
\]
\vspace{-7mm}
\begin{equation}
\label{58}
\end{equation}
\vspace{-7mm}
\[
\left[\psi_{[m4]}(\textbf{x},t),\psi_n(\textbf{y},t) \right]=
\delta_{mn}\delta(\textbf{x}-\textbf{y}),~~~~
\left[\psi_{[mn]}(\textbf{x},t),\widetilde{\psi}_k(\textbf{y},t)
\right]= \epsilon_{mnk}\delta(\textbf{x}-\textbf{y}),
\]
where $\epsilon_{mnk}$ is antisymmetric tensor
($\epsilon_{123}=1$).

The operators for the neutral field, in the second quantized theory,
can be written as follows:
\[
\Psi(x)=\sum_{k,s}\left[a_{k,s}\Psi^{(+)}_{s}(x) +
a^+_{k,s}\Psi^{(-)}_{s}(x)\right] ,
\]
\vspace{-7mm}
\begin{equation} \label{59}
\end{equation}
\vspace{-7mm}
\[
\overline{\Psi}(x)=\sum_{k,s}\left[a^+_{k,s}
\overline{\Psi}^{(+)}_{s}(x)+ a_{k,s}
\overline{\Psi}^{(-)}_{s}(x)\right] ,
\]
where the positive and negative parts of the wave function are
given by (24). The creation and annihilation operators of
particles, $a^+_{k,s}$, $a_{k,s}$, obey the commutation relations
\cite{Kruglov}, \cite{Kruglov1}:
\begin{equation}
[a_{k,s},a^+_{k',s'}]=\varepsilon_s\delta_{ss'} \delta_{kk'} ,~~~
[a_{k,s},a_{k',s'}]=[a^+_{k,s},a^+_{k',s'}]=0, \label{60}
\end{equation}
where $\varepsilon_s=1$ at $s=\widetilde{0},m$, and
$\varepsilon_s=-1$ at $s=0,\widetilde{m}$. There is no summation
in the index $s$ in (60). The operators $a_{k,s}$, $a^+_{k,s}$
at $s=0,\widetilde{m}$, corresponding to scalar and pseudovector
states, satisfy the commutation relation with the ``wrong"
sign ($-$) and one should introduce the indefinite metric
\cite{Kruglov1}. The energy density follows from (36), and is
\begin{equation}
 {\cal E}=-T_{44}=\pi (x)\partial_0\Psi (x) -{\cal L}
=i\overline{\Psi}(x)\Gamma_4\partial_0\Psi(x). \label{61}
\end{equation}
With the help of (59)-(61), and the normalization conditions,
we obtain the Hamiltonian
\begin{equation}
H=\int {\cal E}d^3 x=\sum_{k,s}k_0\varepsilon_s\left(a^+_{k,s}
a_{k,s}+a_{k,s} a^+_{k,s}\right) . \label{62}
\end{equation}
After the introduction of the indefinite metric the eigenvalues of
the Hamiltonian (62) are positive values but the classical
Hamiltonian is not positive-definite. As a result, scalar and
pseudovector states are ghost states \cite{Kruglov1}.

From (59),(60), one finds commutation relations for different
times:
\begin{equation}
[\Psi_M(x),\Psi_N(x')]=[\overline{\Psi}_{ M}(x), \overline{\Psi}_{
N}(x')] =0, \label{63}
\end{equation}
\begin{equation}
[\Psi_ { M}(x),\overline{\Psi}_{N}(x')]=S_{ MN}(x,x'), \label{64}
\end{equation}
\begin{equation}
S_{MN}(x,x')=S^+_{MN}(x,x')-S^-_{ MN}(x,x') , \label{65}
\end{equation}
\begin{equation}
S^+_{MN}(x,x')=\sum_{k,s}\varepsilon_s\left(\Psi^{(+)}_{s}(x)\right)_M
\left(\overline{\Psi}^{(+)}_{s}(x')\right)_N  ,\label{66}
\end{equation}
\begin{equation}
S^-_{MN}(x,x')=\sum_{k,s}\varepsilon_s\left(\Psi^{(-)}_{s}(x)\right)_M
\left(\overline{\Psi}^{(-)}_{s}(x')\right)_N . \label{67}
\end{equation}
From (24),(64)-(67), we obtain:
\begin{equation}
S^\pm_{MN}(x,x')=\sum_{k,s}\frac{\kappa}{2k_0
V}\varepsilon_s\left(u_{s}(\pm
k)\right)_M\left(\overline{u}_{s}(\pm k)\right)_N\exp [\pm
ik(x-x')] .
 \label{68}
\end{equation}
From (27), and taking into consideration that the sum of all
spin projection operators is unity (see \cite{Kruglov1}), one finds
\begin{equation}
\sum_s\varepsilon_s\left(u_{s}(\pm
k)\right)_M\left(\overline{u}_{s}(\pm k)\right)_N =\left(\kappa
\overline{P}\mp i\hat{k}\right)_{MN}. \label{69}
\end{equation}
With the help of (69), we obtain from (68):
\[
S^\pm_{MN}(x,x')=\sum_{k} \frac{1}{2k_0 V} \left(
\kappa\overline{P}\mp i\hat{k}\right)_{MN} \exp [\pm ik(x-x')]
\]
\vspace{-6mm}
\begin{equation} \label{70}
\end{equation}
\vspace{-6mm}
\[
= \left( \kappa\overline{P}-\Gamma_\mu\frac{\partial}{\partial
x_\mu}\right)_{MN} D_{\pm}(x-y),
\]
where we exploit the singular functions \cite{Ahieser}
\begin{equation}
D_+(x)=\sum_{k}\frac{1}{2k_0V}\exp
(ikx),~~~~D_-(x)=\sum_{k}\frac{1}{2k_0V}\exp (-ikx). \label{71}
\end{equation}
Introducing the function \cite{Ahieser}
\begin{equation}
D_0 (x)=i\left(D_+(x)-D_-(x)\right), \label{72}
\end{equation}
from (65)-(67),(70),  we arrive at
\begin{equation}
S_{ MN}(x,x')=-i\left(
\kappa\overline{P}-\Gamma_\mu\frac{\partial}{\partial
x_\mu}\right)_{MN} D_{0}(x-x').
 \label{73}
\end{equation}

One can prove, with the help of (8),(10),(28), that the
relations $\Gamma_\mu\overline{P}=P\Gamma_\mu$, $P\overline{P}=0$
are valid. Thus, we find the equation
\begin{equation}
 \left(\Gamma_\mu\partial_\mu+ \kappa
P \right)\left(
\kappa\overline{P}-\Gamma_\mu\partial_\mu\right)=-\partial_\alpha^2.
 \label{74}
\end{equation}
With the aid of the singular function properties \cite{Ahieser}, one
can verify the relation
\begin{equation}
\left(\Gamma_\mu\frac{\partial}{\partial x_\mu} +\kappa
P\right)S^\pm(x,x')=0 .
 \label{75}
\end{equation}
The propagator (the vacuum expectation of the chronological pairing of
operators) is defined by the equation
\[
\langle T\Psi_{M}(x)\overline{\Psi}_{ N}(y)\rangle_0=S^c_{
MN}(x-y)
\]
\vspace{-6mm}
\begin{equation} \label{76}
\end{equation}
\vspace{-6mm}
\[
=\theta\left(x_0 -y_0\right)S^+_{MN}(x-y)+\theta\left(y_0
-x_0\right)S^-_{MN}(x-y) ,
\]
where the theta-function is $\theta(x)$. Using the function
$D_c (x-y)$ \cite{Ahieser}:
\begin{equation}
D_c (x-y)=\theta\left(x_0 -y_0\right)D_+(x-y)+\theta\left(y_0
-x_0\right)D_-(x-y) , \label{77}
\end{equation}
we obtain the propagator
\begin{equation}
\langle T\Psi_{M}(x)\overline{\Psi}_{N}(y)\rangle_0 =\left(
\kappa\overline{P}-\Gamma_\mu\frac{\partial}{\partial
x_\mu}\right)_{MN}D_c (x-y).
 \label{78}
\end{equation}

Taking into account the equation \cite{Ahieser}
\[
\partial_\mu^2D_c(x)=i\delta(x),
\]
we arrive at
\begin{equation}
\left(\Gamma_\mu\frac{\partial}{\partial x_\mu}+ \kappa
P\right)\langle T\Psi(x)\cdot\overline{\Psi}(y)\rangle_0 =
-i\delta (x-y).
 \label{79}
\end{equation}
The propagator (79) corresponds to Dirac$-$K\"{a}hler's massless
fields including scalar, vector, pseudovector and pseudoscalar fields.

\section{Discussion}

We have considered the first-order formulation of the theory of
Dirac$-$K\"{a}hler massless fields. The form-invariance
of the first-order relativistic wave equation was proven.
This is a new type of formulation of Dirac$-$K\"{a}hler
antisymmetric tensor fields. This allows us to obtain in
the simple manner the canonical and symmetrical Belinfante
energy-momentum tensors. The traces of the energy-momentum tensors
do not equal zero and the divergences of the canonical and Belinfante
dilatation currents do not vanish. Nevertheless, we obtain a new conserved
dilatation current but the conformal invariance is broken. The canonical
quantization requires the introduction of indefinite metrics. This
is connected with the presence of ghosts (the pseudovector and
scalar states). The propagator of the massless fields in the
matrix form has been found and can be used for some calculations
in the quantum theory.

If we impose the conditions $\psi_0 (x)=0$, $\widetilde{\psi}_0
(x)=0$, we arrive at the two-potential formulation of
electrodynamics which is convenient for considering magnetic
monopoles \cite{Kruglov1}.

Let us discuss the possible important application of the
theory considered with the nonperturbative QCD on a lattice.
First, the Dirac$-$K\"{a}hler formulation of fermions on the lattice
is an interesting problem \cite{Beauce}.
We notice that all tensor fields considered are described by the $16$-component
wave function, i.e. they form the same multiplet. To have a connection
with flavor degrees, one needs to make
the transformation of the matrices and the wave function:
\[
\Gamma'_\mu=S\Gamma_\mu S^{-1}=I\otimes \gamma_\mu,~~~~\Psi'(x)=S\Psi(x)=\{\psi_i\}~~~~(i=1,2,3,4),
\]
where $I$ represents the $4\times 4$ unit matrix, and $\gamma_\mu$ are the Dirac matrices. In this case the matrices $\Gamma'_\mu$ represent
the direct sum of four Dirac matrices. Then, we formally can express four Dirac spinors $\psi_i$
(flavor degrees) through tensor fields. It should be noted, however, that transformations of tensor fields and
Dirac spinors under the Lorentz group are different \cite{Kruglov}, \cite{Kruglov1}.
We also mention that the Dirac$-$K\"{a}hler formulation on a lattice and the staggered fermion formulation
(which is widely used for lattice QCD) are equivalent \cite{Kawamoto}.
There is another possible application of Dirac$-$K\"{a}hler formulation considered.
It has been pointed out that the Dirac$-$K\"{a}hler fermion formalism is essentially equivalent
to the twisting of topological field theory generating SUSY \cite{Kawamoto1}.
The Dirac$-$K\"{a}hler formulation has also a fundamental connection with the regularization of fermions,
and is related to the twisting of supersymmetry and leads to the corresponding lattice SUSY formulation.
Applicability of the current formulation of Dirac$-$K\"{a}hler fields to the above mentioned investigations
will be the subject of further work.

 \vspace{3mm}

 \textbf{Acknowledgement}

 \vspace{3mm}

I am grateful to a referee of EPJC for his valuable remarks.

\end{document}